\documentclass[aps,twocolumn,showpacs]{revtex4}
\usepackage{graphicx}
\usepackage{amsmath}
\newcommand{\be}{\begin{equation}}
\newcommand{\ee}{\end{equation}}

\begin{document}
\begin{sloppy}

\title{Observation of Enhanced Beaming from Photonic Crystal Waveguides}

\author{Steven K. Morrison and Yuri S. Kivshar}

\affiliation{Nonlinear Physics Centre and Centre for Ultrahigh-bandwidth Devices for Optical Systems (CUDOS), Australian National University,
Canberra ACT 0200, Australia}

\begin{abstract}
We report on the experimental observation of the beaming effect in
photonic crystals enhanced via surface modes. We experimentally map
the spatial field distribution of energy emitted from a
subwavelength photonic crystal waveguide into free-space, rendering
with crisp clarity the diffractionless beaming of energy. Our
experimental data agree well with our numerical studies of the
beaming enhancement in photonic crystals with modulated surfaces.
Without loss of generality, we study the beaming effect in a
photonic crystal scaled to microwave frequencies and demonstrate the
technological capacity to deliver long-range, wavelength-scaled
beaming of energy.
\end{abstract}

\pacs{42.70.Qs, 42.60.Jf, 42.25.Fx, 78.68.tm}
\maketitle

Free-space propagation of light is inherently linked to diffraction; especially so when the beam emerges from a subwavelength
aperture~\cite{Bethe:1944-163:PREV}. A consequence of tight confinement of electromagnetic waves is the large angular diffraction experienced by
the waves abruptly released from the strong confinement.  This substantial diffraction produced by subwavelength apertures is contradictory to
the formation of a highly collimated beam. This contradiction can, however, be brought to accord and a narrow beam fashioned through the
coherent coupling of additional slowly radiating surface waves on the portal surface surrounding the diminutive aperture. The radiating surface
waves behave as Huygens-styled sources that combine and interfere constructively and destructively with all the emissions from the aperture to
form a narrowly confined beam that exhibits diffraction free propagation over substantial distances~\cite{Monroe:2004-11:PRF}.

Beaming of light from a subwavelength aperture was first
experimentally demonstrated in metallic thin films flanked by
periodic corrugations, with surface plasmons employed to transform
the emissions into a confined beam of
light~\cite{Ebbesen:1998-667:NAT}. Through insightful comparisons,
beaming from subwavelength waveguides in photonic crystals  was
theoretically predicted~\cite{Moreno:2004-121402:PRB} and
independently observed in experiment~\cite{Kramper:2004-113903:PRL}.
Further research was conducted into the enhancement of the beaming
effect from photonic crystal waveguides through the engineering of
the modulated surfaces of photonic
crystals~\cite{Morrison:2005-81110:APL,Morrison:2005-343:APB}.

Important factors in the attainment of free-space beaming from
photonic crystal waveguides are the efficient coupling from
waveguide mode to the surface modes and the slow coupling and
steering of radiated surface modes, such that their collective
spatial phase and amplitudes superimpose and interfere destructively
everywhere but directly in front of the waveguide
aperture~\cite{Moreno:2004-121402:PRB}. Monatomic photonic crystals
support surface modes through an introduction of a defect structure
to the surface layer. These surface modes occur below the light line
(the free-space continuum of states) and decay evanescently away
from the surface, and like the waveguide mode, occur within the
photonic crystal's bandgap. If the surface defect structure of a
photonic crystal is also periodically modulated or corrugated, these
important factors required for beaming can be
obtained~\cite{Moreno:2004-121402:PRB,Morrison:2005-81110:APL}.

\begin{figure}[b!]
\begin{center}
\includegraphics[width=8.6cm,keepaspectratio=true]{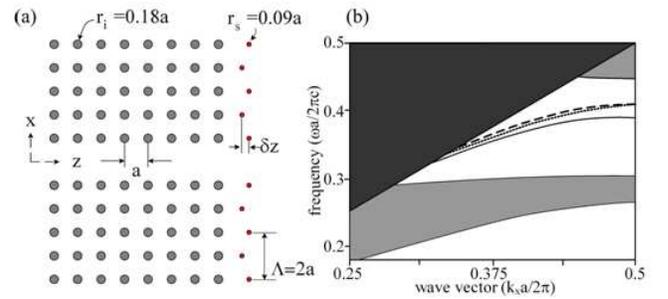}
\caption{(a) Square lattice, two-dimensional photonic crystal
beaming model, illustrating the crystal's geometry, including a
single-row defect waveguide and corrugated mono-defect surface layer
in which surface cylinders have a radius of half the bulk photonic
crystal's cylinder radius. (b) Dispersion relationship of the
photonic crystal depicted in (a) along the direction of the surface
where \emph{even} number cylinders, numbered symmetrically and
sequentially away from the waveguide are moved: (dashed line)
towards the photonic crystal;  (dotted line) positioned in line with
odd number cylinders, removing the surface corrugation; and, (solid
line) away from the photonic crystal. The dark upper triangle
denotes the free-space continuum of states above the light line,
whereas the other shaded regions depict the projected Bloch modes of
the bulk photonic crystal.}
\end{center}
\end{figure}

In this Letter we verify experimentally earlier theoretically
predictions and demonstrate a substantially enhanced beaming of
light from a photonic crystal waveguide of subwavelength width. We
highlight the methods used to improve coupling between the photonic
crystal waveguide mode and surface modes, and illustrate how this
coupling is compatible with setting the correct phase delay and
emission direction of the radiative surface modes to achieve highly
collimated beaming. Furthermore, we demonstrate the substantial
distance over which the beam remains diffractionless and reveal the
Bessel-like beam profile supporting this extensive beaming range.

We commence our investigation into the beaming effect by modeling an archetype photonic crystal using rigorous numerical methods.  The model we
employ is a photonic crystal with a square lattice populated with high dielectric cylinders, with dielectric constant of $\epsilon_{\emph{r}} =
11.54$ (equal to the dielectric constant of alumina at $\sim 10GHz$, as used in the experiment confirmation) and of normalized radii $r=0.18 a$,
where $a$ is the lattice period. Photonic crystals of this configuration produce in-plane band-gaps in the normalized frequency band of $\omega
= 0.3$ to $0.42 \times 2\pi c/a$ when the incident field is polarized with the electric field parallel to the axis of the cylinders. Through the
removal of a single row of cylinders perpendicular to the surface, a single mode waveguide is created; whereas the defect surface structure used
to induce a surface mode is composed of cylinders with a radius reduced by half. To co-join the surface modes and the directly transmitted
field, a surface corrugation is formed by displacing alternative surface defect cylinders away from their lattice sites on the portal defect
surface of the photonic crystal. The important geometric features of the photonic crystal model are illustrated in Fig.~1(a).

To reveal the critical properties of the photonic crystal model, we
calculate the dispersion relationships of the quasi-periodic,
corrugated surface structures using an \emph{ab initio}
finite-difference time-domain (FDTD) method. The method represents
the fields in complex notation and varies the phase relationship
$\exp (i\textbf{k}\textbf{r})$ at the periodic boundaries and across
the one-dimensional Brillouin zone. The width of the super-cell used
is $2a$, such that it includes the double-periodicity of the surface
corrugation. To remedy band folding that occurs due to the doubled
super-cell width, sources are introduced into every unit cell of the
super-cell with phase values determined by Bloch theorem.  Using
this method, we determine that a red shift occurs in the surface
mode frequency for a positive displacement (away from the crystal)
of the corrugation forming cylinders, and conversely, a minor blue
frequency shift occurs for negatively displaced corrugation forming
cylinders, as shown in Fig.~1(b).

Next, we study the important issue of coupling to the surface modes. Again, using the FDTD method we measure the power coupled into the surface
mode, under steady state and numerically converged conditions, as we alter the position of the surface cylinders closer to the waveguide.
Movement of these surface cylinders produces maximum coupling for a forward displacement (away from the crystal). This forward displacement
creates both a cavity-like mode that captures a large quantity of the field, as well as delivering a strong re-scattering of the surface mode
from the subsequent surface cylinders in a direction close to the waveguide aperture, enabling strong initial shaping of the transmitted beam.

To achieve optimal beaming we note the need to achieve both strong coupling to the surface modes along with directional control and phase
alignment of the re-scattered surface modes. From the numerical analysis of the serial coupling to the surface modes, we implement a positive
surface corrugation. From the dispersion analysis, we control the spatial phase positioning of the radiative surface modes by altering the
corrugation displacement in accordance with the results given by Fig.~1(b). This corrugation displacement is, however, tempered and fine-tuned
by the radius of the surface defect cylinders that, in concert with the corrugation displacement, also contribute to the scattering direction of
the leaky surface mode, and \emph{ipso facto}, the operational frequency and group velocities of the surface modes. Optimal displacement of the
surface corrugation cylinders for this photonic crystal is $0.4a$ away from their lattice sites. Figure 2 demonstrates the strong contrast
between (a) the simple terminated photonic crystal with full-size surface cylinders and (b) the optimal beaming, occurring at $\omega = 0.378
\times 2\pi c/a$, with half-diameter surface cylinders, alternatively displaced positively by $0.4a$, starting from the waveguide aperture and
with a periodicity of $2a$. From these results, we determine that $82\%$ of the transmitted field is conveyed into the narrow directed emission
that in comparison to the un-altered surface yields a $258\%$ increase in power within the area of the centrally directed beam.

\begin{figure}[t!]
\begin{center}
\includegraphics[width=8.6cm,keepaspectratio=true]{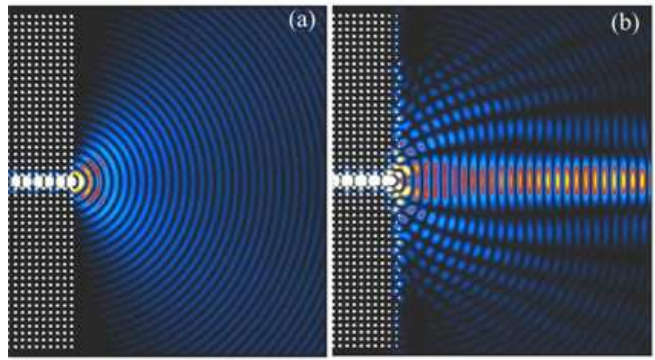}
\caption{Finite-difference time-domain simulations of the field
intensity from the beaming modeling with (a) illustrating the full
angle diffused output from the sub-wavelength waveguide exit when
the surface structure is simply terminated in a row of cylinders
equal in radius to those of the bulk photonic crystal, and (b)
illustrating the highly collimated beam emitted from the waveguide
having been strongly fashioned by radiative surface states.}
\end{center}
\end{figure}

\begin{figure}[t!]
\begin{center}
\includegraphics[width=8.6cm, keepaspectratio=true]{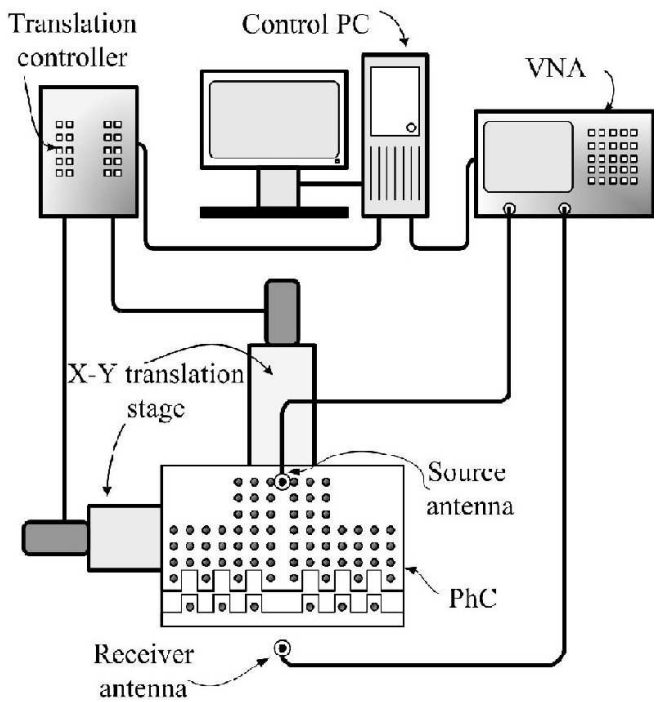}
\caption{ Schematic representation of the pertinent elements and
experimental configuration used to map the field intensity of the
beaming effect. }
\end{center}
\end{figure}

Experimental fabrication of this photonic crystal for operation at optical wavelengths is very challenging, requiring significant fabrication
resources. In order to confirm enhanced beaming, yet avoid these fabrication complexities, we scale the crystal to microwave wavelengths
centered around 10GHz, where the lattice period is 11mm and use alumina cylinders with a refractive index of $\sim 3.4$. Figure 3 shows a
schematic of the experimental setup with broadband source and detection facilitated by a 20GHz vector network analyzer (Rohde and Schwartz
ZVB20) interfaced to half-wavelength dipole antennas. To achieve a low impedance transition from the unbalanced coaxial cable connected from the
vector network analyzer (the conductors of the coax do not connect to the antenna in the same way and are thus unbalanced) to the symmetrical,
balanced dipole antenna, a miniature sleeve balun with a $1:1$ impedance transform was used. The source dipole antenna inserted into the
crystal's waveguide excites a waveguide mode with an electric field parallel to both the antenna and cylinder axis after approximately four
periods along the waveguide, as confirmed by FDTD simulations. Electromagnetic energy not coupled into the waveguide mode exits the rear of the
waveguide. We mount the detector antenna to a tripod in front of the photonic crystal. Finally, to map the transmitted field, we mount the
photonic crystals to an X-Y translation stage and use a phase maintaining cable between the source dipole antenna and the network analyzer, thus
removing any phase noise occurring due to the disturbance of the cabling as the translation stage moves while mapping the transmitted field.

\begin{figure}[b!]
\begin{center}
\includegraphics[width=8.6cm,keepaspectratio=true]{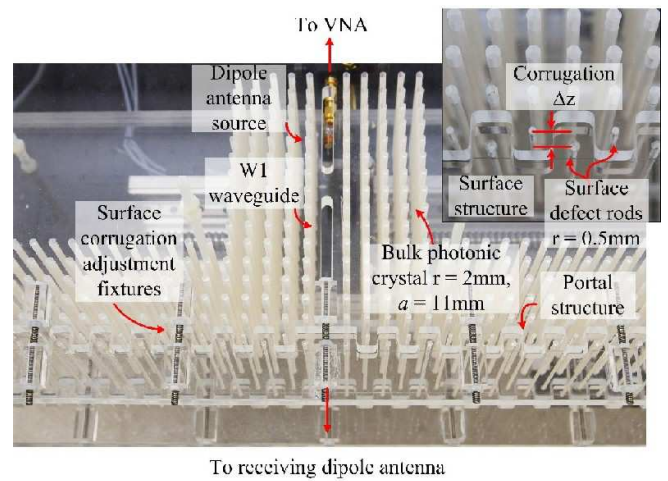}
\caption{Photograph of the experimental photonic crystal and portal
surface. (Inset) Close-up of the photonic crystal's defect portal
surface, highlighting the surface corrugation that, due to momentum
conservation of the surface states, allows the surface modes to
radiate into free space. }
\end{center}
\end{figure}

\begin{figure*}
\begin{center}
\includegraphics[width=15cm,keepaspectratio=true]{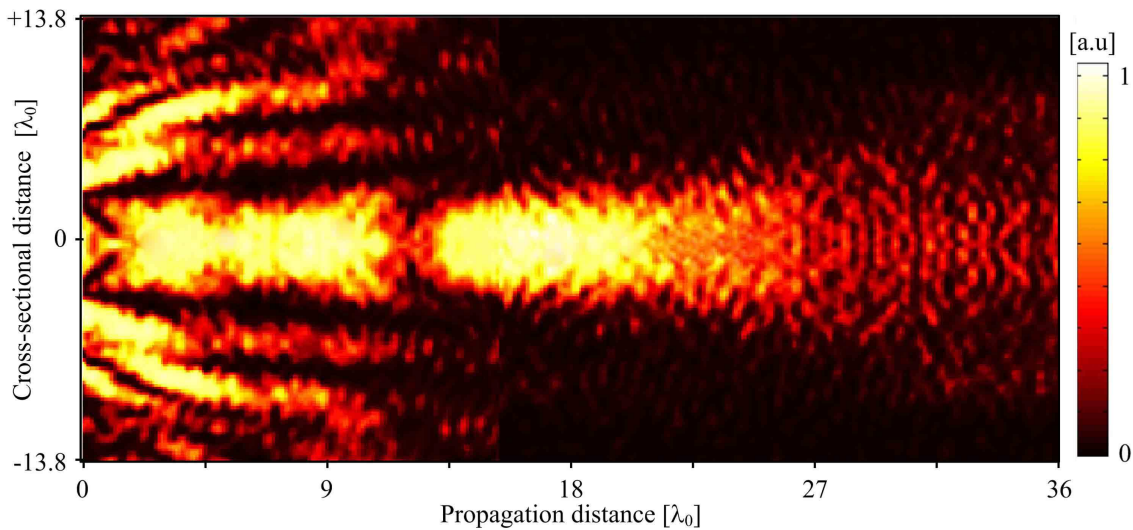}
\caption{Experimental map of the intensity field from the photonic
crystal waveguide demonstrating the beaming effect at 10.4GHz in a
plane central to the height of the crystal's cylinders and in front
of the crystal's portal surface (spatial dimensions are given in
free space wavelengths equivalent to 28.82mm). }
\end{center}
\end{figure*}

Figure 4 shows the surface structure of the crystal formed with 1mm radius cylinders with a corrugation period of $2a$. The surface corrugation
displacement, $\Delta z$, is slide adjustable as illustrated in the inset of Fig. 4. This adjustable corrugation displacement allows adjustment
and fine-tuning of the beaming effect to overcome experimental discrepancies in, for example, the dielectric constant of the alumina cylinders.
We sample the field on a 5mm grid, equating to approximately 6 points per wavelength, and then interpolate the measured result using a
two-dimensional cubic spline; Fig. 5 illustrates the results at 10.4GHz with a corrugation displacement of 4.44mm. Owing to impedance-mismatch
return-losses between the $50 \Omega$ vector network analyzer, antennas and Bloch impedance of the photonic crystal's waveguide, the field is
plotted in linear, normalized arbitrary units as indicated by the color scale.

\begin{figure}[htb!]
\begin{center}
\includegraphics[width=8.3cm,keepaspectratio=true]{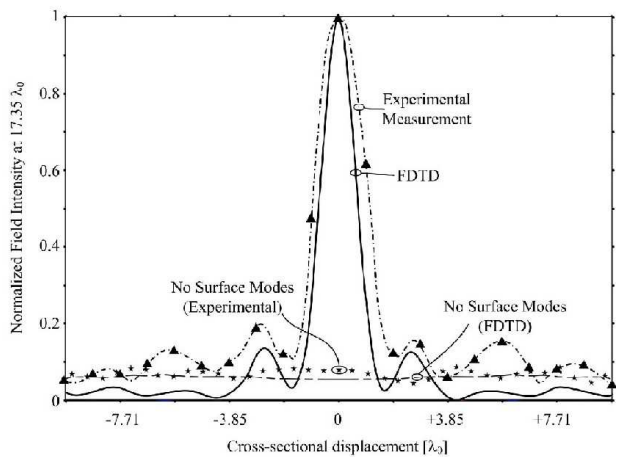}
\caption{Experimentally and numerically determined cross sections of
the normalized field intensity at a distance of 17 $\lambda_0$ in
front of the portal surface of the photonic crystal, demonstrating
both virtually flat field distribution in the absence of radiating
surface modes and highly collimated transmission, with Bessel-like
profile, resulting from the beaming effect.}
\end{center}
\end{figure}

Clearly shown in Fig. 5 is the crisp clarity of the mapped intensity field, revealing the long distance over which anomalously low divergence of
the transmitted beam occurs, as well as demonstrating a strong correlation to the numerical result. Also clearly visible in Fig. 5 are the side
lobes resulting from the grating effect of the surface corrugation, thus clearly confirming the theory that leaky surface modes shape the
outgoing emission and transform the normally diffused output into an anomalously low diverging beam. Through mapping of the field without
corrugation, we observer and affirm a strong, rapidly diffracting emission, absence of beaming. It is worth commenting that the long range
diminishing of the experimental directed emission is to a large degree a result of the diffraction of the beam in the non-confined, out-of-plane
direction resulting from the two-dimensional nature of the photonic crystal. Furthermore, the finite height of the experimental crystal
($~5\lambda$) that is bounded by low dielectric constant Perspex plates causes spurious, destructive interference of the beam at approximately
20 wavelengths in front of the crystal's surface.

Cross-sectional slices of the normalized field intensity for both the numerical and experimental results are taken at a distance of $17
\lambda_0$ in front of the portal surface of the photonic crystals and are presented in Fig. 6, with linear normalization taken with reference
to the central intensity peak of the numerical results.  Also clearly seen in Fig 6. is the Bessel-like beam profile, known to propagate without
diffraction over substantial distances, formed by the leaky surface modes.

In conclusion, we have studied, theoretically and experimentally, the conditions for dramatic enhancement of the free-space beaming of
electromagnetic energy from a subwavelength photonic crystal waveguide. These conditions, that have a tendency to contend with each other, can
be brought to accord such that optimal superposition of all emissions, both direct and re-scattered from the surface of the crystal, produce a
highly collimated beam of wavelength dimensions. We have clearly demonstrated, with a strong correlation between our theoretical and
experimental results, the role of the radiative surface modes in the fashioning of the directed emission with a Bessel-like beam profile capable
of propagation over substantial distance without divergence. Our results hold great technological promise and perspectives for tailoring the
beam profiles from subwavelength apertures in microwave, millimeter wave and optical systems.

This work has been supported by the Australian Research Council through the Discovery and Center of Excellence projects. We thank Dr. Ilya
Shadrivov for useful discussions and help with experiments.

\end{sloppy}
\end{document}